\documentclass[%
 reprint,
 amsmath,amssymb,
 aps,
 prl,
 floatfix,
]{revtex4-1}

\usepackage[utf8]{inputenc}
\usepackage{braket}
\usepackage{graphicx}
\usepackage{dcolumn}
\usepackage{bm}
\usepackage[normalem]{ulem} 
\usepackage{color}

\usepackage{hyperref}
\hypersetup{
    colorlinks,%
    citecolor=blue,%
    linkcolor=blue,%
    urlcolor=blue
}

\begin{document}

\title{High-degeneracy points protected by site-permutation symmetries}

\author{F. Crasto de Lima}
\email{felipe.lima@ufu.br}
\author{G. J. Ferreira}
\affiliation{Instituto de F{\'i}sica, Universidade Federal de Uberl{\^a}ndia, \\
       C.P. 593, 38400-902, Uberl{\^a}ndia, MG,  Brazil}%
\date{\today}

\begin{abstract}
Space-group symmetries dictate the energy degeneracy of quasiparticles (e.g., electronic, photonic) in crystalline structures. For spinless systems, there can only be double or triple degeneracies protected by these symmetries, while other degeneracies are usually taken as \textit{accidental}. In this Rapid Communication we show that it is possible to design higher degeneracies exploring site permutation symmetries. These design principles are shown to be satisfied in previously studied lattices, and new structures are proposed with three, four, and five degeneracy points for spinless systems. The results are general and apply to different quasiparticle models. Here, we focus on a tight-binding approach for the electronic case as a proof of principle. The resulting high-degeneracy points are protected by the site-permutation symmetries, yielding pseudospin-1 and -2 Dirac fermions. The strategy proposed here can be used to design lattices with high-degeneracy points in electronic (e.g., metal-organic frameworks), photonic, phononic, magnonic, and cold-atom systems.
\end{abstract}

\maketitle


Condensed matter systems are not constrained by the Poincaré symmetry. Therefore a rich variety of quasiparticles, without high-energy counterparts, emerge \cite{SCIENCEBradlyn2016}. For instance, higher pseudospin Dirac and Weyl quasiparticles have been studied in Refs.~\cite{PRBDora2011, PRBLan2011, PLAFeng2019}. These quasiparticles, arising in high-degeneracy points, present enhanced Klein tunneling \cite{PRBBetancur2017}, and superscattering regimes \cite{PRAXu2017}. Moreover, associated with high-degeneracy (HD) points, hourglass \cite{NATUREWang2016, SCIENCEMa2017, PRLWang2019} and Kane \cite{NATUREOrlita2014, JPCLWang2018} fermions have been proposed and observed in three-dimensional (3D) and two-dimensional (2D) systems. Focusing on technological applications, beyond the new quasiparticles, a synergy between HD points in the electronic and phononic energy spectra allow for an enhanced thermoelectric performance \cite{NATURESnyder2008, AEMFu2014}. However, the space group symmetries, for spinless systems, only constrain the states to have at most twofold (in 2D systems) and threefold (in 3D systems) degeneracy, apart from \textit{accidental} degeneracies \cite{tinkham, Dresselhaus}. Finding degeneracy points beyond the space-group symmetries constraints is a key to the emergence of new topological and semimetal phases \cite{NATURELu2014, PRBZhu2018, NATUREGao2018}.

Currently, an extensive case by case search for HD points has been the object of different studies \cite{SCIENCEBradlyn2016, NATURESchroter2019, cano2019multifold}. However, a design principle to achieve such HD points is still missing. Concomitantly, we are facing great experimental control of lattice formations \cite{OPTLETTFelix2008, OPTICABryn2017, PRLCerjan2019}, where threefold band degeneracy has been experimentally observed in a photonic \cite{NATUREYang2019} and in cold-atom \cite{PRBShen2010} lattices. Additionally, state-of-the-art organic chemistry allows for a combination of differently designed molecules, yielding nontrivial structures \cite{Colson2013, CCRodriguez2016, NATUREBaumann2019}.

In this Rapid Communication, we analyze symmetry-protected HD occurring in periodic lattices. For an $N$ equivalent site spinless lattice, we find that $(N-1)$-degenerated states are protected by a set of $(N-1)$ site-permutation symmetries. This allows us to propose {weak and strong constraints as} a design principle for these HD points, which we illustrate for a few 2D and 3D lattices. 
{Within the weak constraint regime, the HD is guaranteed for models with only first-neighbor hoppings, while lattices that satisfy our strong constraint show HD robust even for long-range interactions. Here we show examples of both cases.}
The generality of the presented discussion allows for an interpretation in the context of different quasiparticle systems: electronic, photonic, phononic, magnonic, and cold-atom \cite{RMPOzawa2019, RMPGu2018, LTPRychly2015, RMPCooper2019}. Here, we focus on electronic spinless tight-binding (TB) models at the $\Gamma$ point ($\bm{k} = 0$) to show that pseudospin-1 and -2 Dirac fermions emerge in the HD points.

\paragraph{Site-permutation symmetries and degeneracy.} In 2D spinless systems with azimuthally symmetric intersite interactions (e.g., $s$, $p_z$, $d_{z^2}$ orbitals in electronic systems), and in 3D for spherically symmetric interactions (i.e., $s$ orbitals in electronic systems), all point-group (PG) operations at $\Gamma$ can be decomposed into combinations of the site-permutation operations \cite{JPCHougen1986}. More interestingly, beyond the PG symmetries, extra site-permutation symmetries can introduce new degeneracies in the Hamiltonian spectra. These can be understood as a consequence of sublattice equivalences.

Let us start with a simple abstract lattice, where all sites are equivalent. Later we will properly define more general conditions for the lattices. First, let $P^{(ij)}$ be a site-permutation operation that exchanges the $i$th and $j$th sites, i.e., a pair permutation \cite{MPLonguet1963}. Considering a basis localized on each site, the matrix representation $D(P^{(ij)})$ of $P^{(ij)}$ has elements $D_{kk}(P^{(ij)}) = 1$ for $k \neq \{i,\,j\}$, $D_{ij}(P^{(ij)}) = D_{ji}(P^{(ij)}) = 1$, and $D_{k\ell}(P^{(ij)}) = 0$ otherwise. Therefore, such operators are real and unitary, $P^{(ij)} = P^{(ji)} = \left[P^{(ij)} \right]^{-1}$. Note that, for a system with $N$ sites, there exist $(N-1)N/2$ possible permutations of pairs. All these permutation symmetries can be decomposed in $N-1$ generators formed, for instance, by the symmetries $P^{(kj)}$ with $k$ fixed and $j\neq k$. Every other pair permutation can be constructed as $P^{(ij)} = P^{(ki)}P^{(kj)}P^{(ki)}$, which form a conjugacy class of the symmetric group $S_N$ \cite{bruce}.

Next, the goal is to find Hamiltonians ($H_{N}$) that commute with all these $N-1$ permutation symmetries. Naturally, if $H_N$ is a sum of the permutation symmetries,
\begin{equation}
    H_N \propto \sum_{i>j}^N D(P^{(ij)}),
\end{equation}
it commutes with all $P^{(ij)}$, since $[P^{(ij)},\,P^{(ik)}] = - [P^{(ij)},\,P^{(kj)}]$, and $[P^{(ij)},\,P^{(kl)}] = 0$ for $\{i,\,j\} \neq \{k,\,l\}$. Within the basis set presented above, $H_N$ becomes
\begin{equation}
H_N = t\,\mathbb{J}_N - (t+m)\mathbb{I}_N = 
\left(
\begin{matrix}
m & t & \dots & t \\
t & m & \dots & t \\
\vdots & \vdots & \ddots & \vdots \\
t & t & \dots & m 
\end{matrix}
\right),\label{hamilt}
\end{equation}
where $t$ is a real number, $\mathbb{J}_N$ is the $N \times N$ all-ones matrix, and $\mathbb{I}_N$ is the $N \times N$ identity matrix. The role of the $m$ number will be further discussed below. For now, note that $m$ just rigidly shifts the eigenvalues of $H_N$. Since the site-permutation symmetries act exchanging lines and columns of the Hamiltonian matrix, Eq. \eqref{hamilt} is the most general form allowed for $H_N$ that is invariant under the permutations, i.e., $P^{(ij)}H_N P^{(ij)} = H_N$. This Hamiltonian has only two sets of eigenvalues ($\lambda_n$) \cite{MatrxAnalysis}, an $N-1$ degenerated set with $\lambda_1 = \lambda_2 = \cdots \lambda_{N-1} = -t+m$, and a nondegenerated $\lambda_N = (N-1)t + m$. Therefore, for $N$ site lattices described by these symmetries, an $N-1$ degeneracy point emerges.

\paragraph{Lattice constraints.} The site-permutation operations $P^{(ij)}$ that commute with $H_N$ are limited to equivalent sites $i \equiv j$. The equivalence here refers to sites with the same local energy, and that couple in the same way with every other site \footnote{{Note that the permutation operation $P^{(ij)}$ exchange the $i$th and $j$th lines/columns of the Hamiltonian. Therefore, in order for $P^{(ij)}H_{N}P^{(ij)} = H_N$ to be fulfilled, the coupling of an $i$th site to any other $p$th site should be equal to the $j$th and $p$th coupling.}}, creating a uniformity in $H_N$. Within a first-nearest neighbor (1NN) {and spherically symmetric} interaction, this uniformity reflects in the coordination number of the sites $i$ and $j$ being the same. This site coordination number can be cast as $c_1=n_1(N-1) + m_1$, where $n_1$ is the number of times a site $i$ couples with each of the remaining $N-1$ sites $j \neq i$ within the same or neighboring cells, and {$m_1$ is the number of times a site $i$ couples with a translated version of itself on neighboring cells.} {This criteria defines a weak constraint, i.e., the minimum ingredient for the arising of the HD points.} This condition can be extrapolated for long-range interactions. For each distance defining a $p$th-nearest neighbor (pNN) interaction, a $c_p=n_p(N-1) + m_p$ coordination condition should be satisfied{, defining a strong constraint into the lattice design}. {If only the weak constraint is satisfied, the degeneracy will strictly occur only for 1NN models, while a small gap is allowed if long-range interactions are included. On the other hand, if the strong constraint is satisfied, the $N-1$ degeneracy is robust.}

\begin{figure}[t]
\includegraphics[width=0.85\columnwidth]{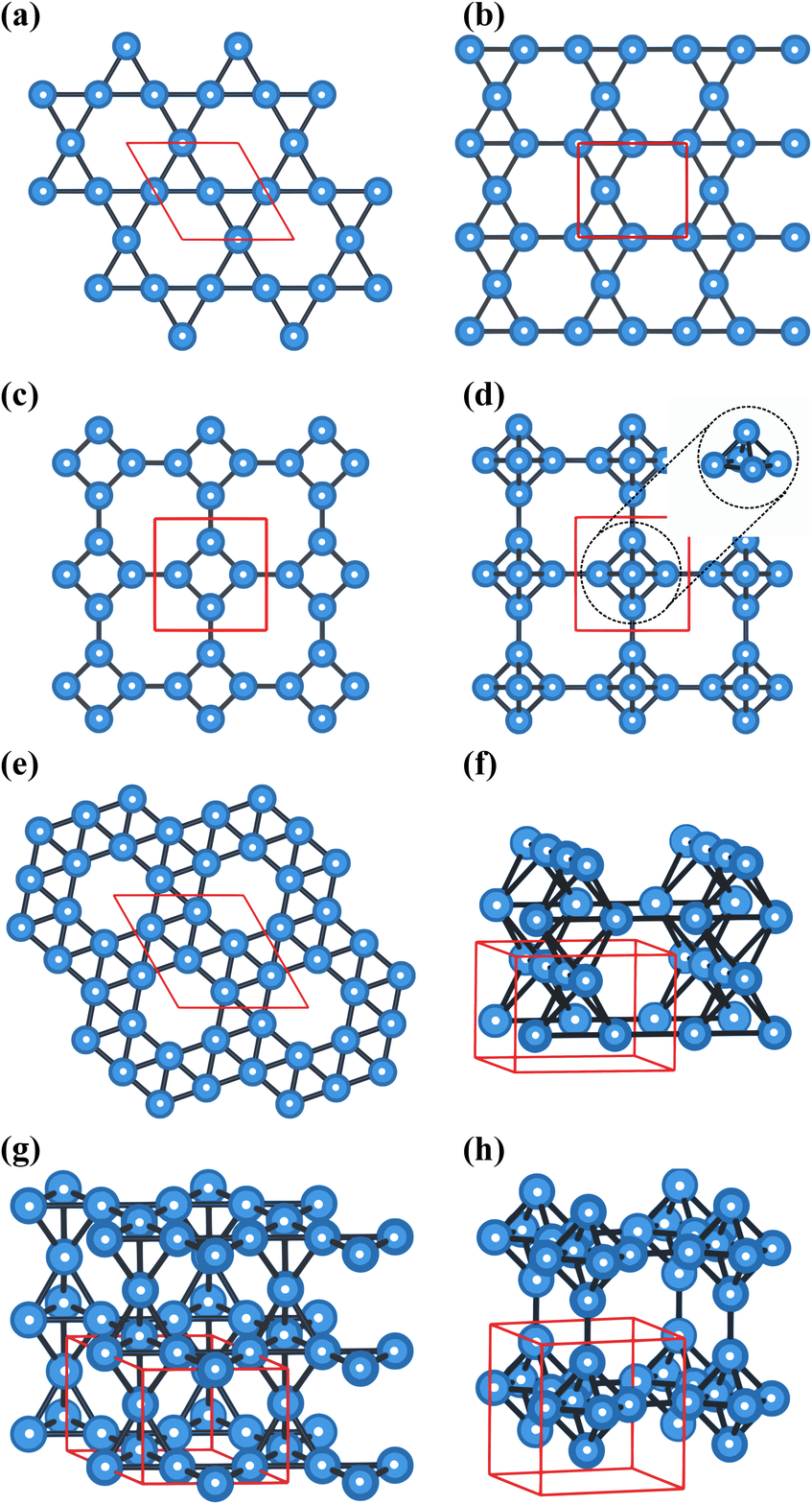}
\caption{\label{lattices} Example of lattices fulfilling the minimum ingredients for HD points{, that is, the $c_1$ coordination criteria}, namely (a) kagome, (b) triangular-rectangular, (c) square-octagonal, (d) pyramidal, (e) snub-hexagonal, (f) crossed-tetrahedral, (g) pyrochlore and (h) cubic-octahedra lattices. {The blue circles indicate the sites positions, black lines connect first neighbors, and red lines define the unit cell.}}
\end{figure}

{In order to find lattices fulfilling the $H_N$ Hamiltonian [Eq. \eqref{hamilt}], we note that $H_N$ takes the form of a graph theory adjacency matrix. In this square matrix the out-of-diagonal terms indicate the connection between sites, which can be directly represented in a real space connected graph \cite{graphbook}. Such graph can then be disentangled into the desired periodic lattices \cite{supmat}.} Below, lattices fulfilling the conditions for at least 1NN interactions {(weak constraint)} are presented (see Fig. \ref{lattices}). Within these lattices, {two cases, Figs. \ref{lattices}(a) and \ref{lattices}(e), also satisfy the long-range condition pNN for $p\rightarrow \infty$ (strong constraint).}
 
\paragraph{Illustrative systems.} Here it is proposed five 2D and three 3D lattices satisfying  at least the 1NN constraint. The lattices shown in Fig.~\ref{lattices} have $N=3,\,4,\,5$ and $6$ equivalent sites within their unit cells, allowing from twofold to fivefold degeneracy. All the proposed lattices have $m_1=0$ within the 1NN constraint. The two-dimensional kagome [Fig. \ref{lattices}(a)] and triangular-rectangular lattices [Fig.\,\ref{lattices}(b)] have $N=3$ and $n_1=2$, i.e., each of the $i=1,2$ and $3$ sites couples two times with the other two {($j \neq i$), as indicated by the black lines connecting the nearest neighbors}. For the square-octagonal, pyramidal, and snub-hexagonal lattices [Figs. \ref{lattices}(c)--\ref{lattices}(e)], with $N=4$, $5$, and $6$, respectively, the $i$th site couples only once with each $j \neq i$ neighbor, thus $n_1=1$. The three-dimensional tetragonal-crossed, pyrochlore, and octrahedral lattices [Figs. \ref{lattices}(f)--\ref{lattices}(h)], have $N=4$, $4$, and $6$, and analogously $n_1=2$, $2$, and $1$, respectively. Interestingly, three of these lattices -- the kagome the square-octagonal, and the snub-hexagonal -- belong to the same class of Archimedean lattices \cite{PCCPCrasto2019}. Additionally, graphene's $p_z$ orbitals satisfy the Hamiltonian $H_2$ within 1NN with $c_1 = 3$ and $m_1=0$, but since $N=2$, there are only nondegenerated states ($N-1 = 1$ degeneracy), excluding the spin degeneracy.

\begin{figure}
\includegraphics[width=\columnwidth]{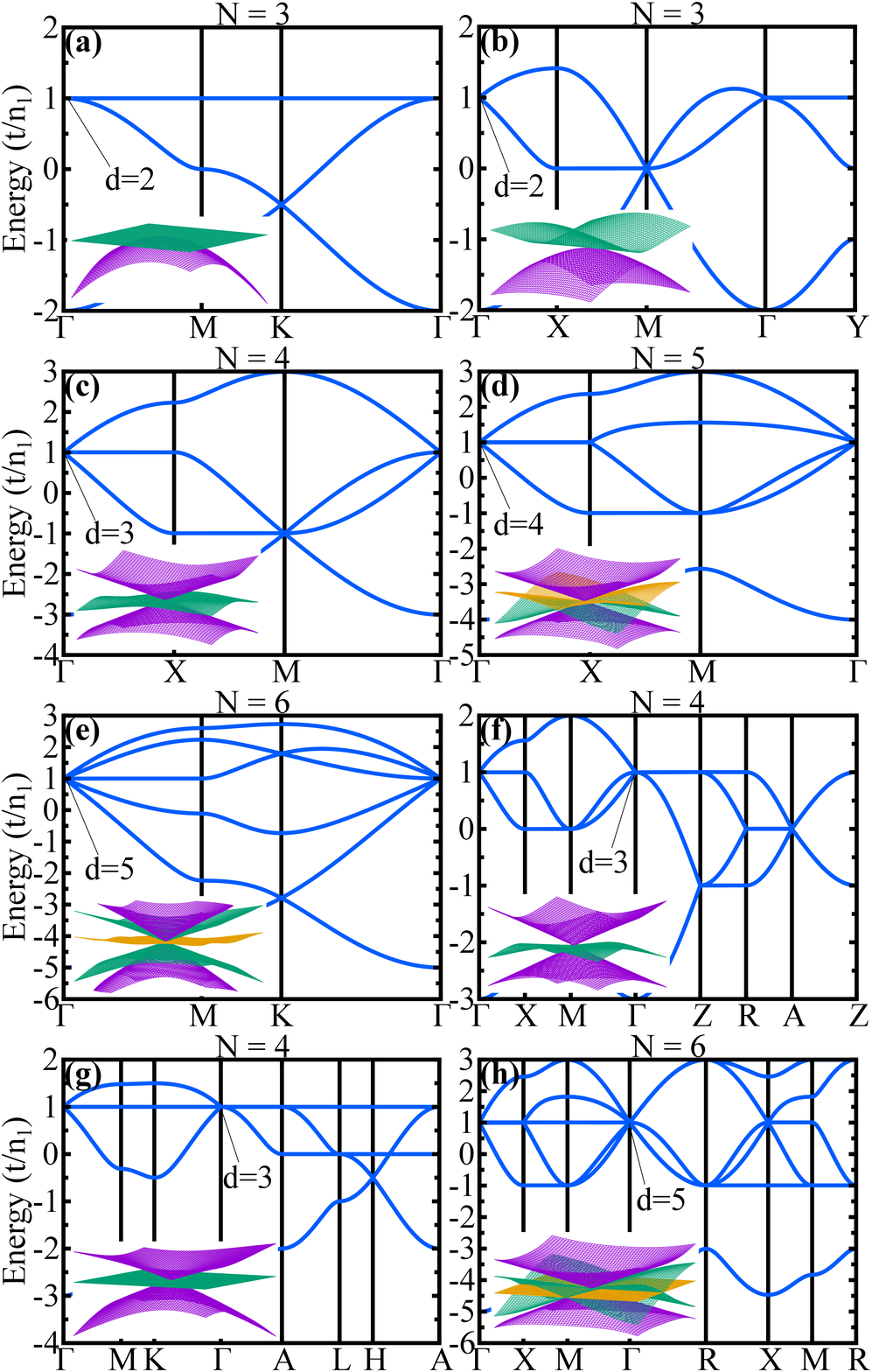}
\caption{\label{tb-model} TB model{, within 1NN interaction ($\alpha=30/d_{\rm 1NN}$),}  for the (a) kagome, (b) triangular-rectangular, (c) square-octagonal, (d) pyramidal, (e) snub-hexagonal, (f) crossed-tetrahedral, (g) pyrochlore and (h) cubic-octahedra lattices. The energy scale is $t/n_1$, where the 1NN hopping strength is $t=-1$, and $n_1$ is the number of equivalent NN}
\end{figure}

These lattices can be designed in photonic \cite{RMPOzawa2019} and cold-atom \cite{RMPCooper2019} systems. Moreover, these proposed lattices have been experimentally observed or theoretically predicted in crystalline materials. For instance, the kagome lattice occurs in metal-organic frameworks (MOFs) \cite{JACSKambe2013, PCCPdeLima2018} and in Fe$_3$Sn$_2$ \cite{NATUREYe2018}, while its 3D version, the pyrochlore lattice, occurs in many ternary oxide systems \cite{RMPGardner2010}. The snub-hexagonal lattice can be grown with MOF surface self-assembly \cite{JACSLiu2011, FDYan2017}, and in a stable boron 2D allotrope \cite{JPCLYi2017, NLCrasto2019}. Still within borophene polymorphs, a structure similar to the triangular-rectangular lattices has been experimentally found \cite{NATURELiu2019}, while a square-octagonal lattice has been shown in a carbon 2D allotrope \cite{PCCPCrasto2019}.

\paragraph{Tight-binding model.} Up until this point no considerations regarding the nature of the $H_N$ were discussed. The analysis above is general for all quasiparticles localized in lattice sites. In order to show the degeneracy points, and its robustness with the site-permutation symmetries, we consider a spinless electronic TB model for each lattice with the on-site energy $\varepsilon_i$ and hopping terms $t_{ij}$ as
\begin{equation}
H_{TB} = \sum_{i} \varepsilon_i c_i^{\dagger}c_i + \sum_{i \neq j} t_{ij}c_i^{\dagger}c_j.
\end{equation}
Unless specified, we assume $\varepsilon_i=0$, and a hopping strength decaying with the distance between the sites as $t_{ij} = t \exp \left( - \alpha d_{ij} \right)$, with $t=\exp \left(\alpha d_{\rm 1NN} \right) =-1$ defining the energy scale, and $d_{ij}$ and $d_{\rm 1NN}$ as the intersite distance and 1NN distance, respectively. The $\alpha$ factor controls the range of the intersite couplings. For $\alpha \gg (d_{\rm 1NN})^{-1}$ only 1NN hoppings are significant to $H_{TB}$, while for $\alpha \ll (d_{\rm 1NN})^{-1}$ a long-range interaction emerges with significant contribution from further neighbors. {Note that this tight-binding approach captures the $H_N$ Hamiltonian, where the diagonal $m$ [Eq. \eqref{hamilt}] values are represented by the on-site energy term, while $t$ [Eq. \eqref{hamilt}] is the hopping term.}

For the kagome lattice ($N=3$), the energy dispersion shows a twofold degeneracy between a flat and a parabolic band at $\Gamma$, Fig. \ref{tb-model}(a), and the third nondegenerate band completes the set of $N$ states from $H_N$. For the triangular-rectangular lattice, the same degeneracies of the kagome lattice occur at $\Gamma$ [Fig. \ref{tb-model}(b)]. However, in this case a flat band is seen only along the $\Gamma$-$Y$ direction, while asymmetrically, a linear Dirac-like dispersion is observed for the $\Gamma$-$X$ and $\Gamma$-$M$ directions. Triple-degeneracy, forming an isotropic pseudospin-1 Dirac dispersion, is observed in the square-octagonal lattice ($N=4$), Fig.\,\ref{tb-model}(c). For the $N=5$ pyramidal lattice of Fig.\,\ref{tb-model}(d), an anisotropic behavior is observed, where the fourfold degeneracy has a Dirac cone with twofold degenerated flat band dispersion along the $\Gamma$-$X$ direction, while two Dirac dispersions with different velocities are present along $\Gamma$-$M$. For the $N=6$ snub-hexagonal lattice the fivefold degeneracy at $\Gamma$ behaves as an isotropic pseudospin-2 Dirac quasiparticle, with the two Dirac cones with velocities distinct by a factor of $2$ [Fig. \ref{tb-model}(e)], and a nearly flat band in between. In the $N=4$ crossed-tetrahedral and pyrochlore 3D lattices [Figs. \ref{tb-model}(f) and \ref{tb-model}(g)] the HD point behaves as a pseudospin-1 Dirac quasiparticle at the $k_z=0$ plane of the Brillouin zone, while in the $k_z$ direction the triple point is formed by two flat bands and a quadratic dispersive one. Lastly, in the $N=6$ cubic-octahedra lattice, the HD point is highly anisotropic, with a Dirac cone and threefold degenerated flat band dispersion for the $\Gamma$-$X$ direction, and doubly degenerated Dirac cones with a {quadratic dispersive} band for the $\Gamma$-$R$ direction. Furthermore, in its $\Gamma$-$M$ direction a pseudospin-2 Dirac-like quasiparticle emerges. 

The presence of these HDs, with the exception of the kagome lattice, could not be predicted from the space-group symmetries of each lattice, since the permutation operations enclose more symmetries than the ones present in the space group. For instance, the space-group symmetries predict at most (i) a nondegeneracy for triangular-rectangular lattice $Pmmm$ space group, while here it was found a twofold degeneracy; (ii) a twofold degeneracy for the kagome ($P6/mmm$), square-octagonal($P4/mmm$), pyramidal ($P4/mm$), snub-hexagonal ($P6/m$), crossed-tetrahedral ($P4_{2}/mmc$), and pyrochlore ($P\bar{6}m2$) lattices, while here it was found from threefold to fivefold degeneracies; and (iii) a threefold degeneracy for the cubic-octahedra $Pm\bar{3}m$ space group, while a fivefold degeneracy was found \cite{bilbao}. {It is worth pointing out that other degeneracy points, not predicted by the space-group, have been found away from $\Gamma$ point. For instance, at the $M$ and $A$ points of the triangular-rectangular and crossed-tetrahedral lattices. Particularly, these degeneracies occur due to interference between the intracell and intercell couplings, with the out-of-diagonal terms vanishing, which leads to a diagonal Hamiltonian with all eigenvalues equal to the on-site energy ($\varepsilon_i = 0$).}

\paragraph{Long-range couplings pNN.} Although lattices can be designed to have only local couplings by tuning the distance between the sites, for instance in photonic lattice constructions \cite{OPTLETTFelix2008, OPTICABryn2017} and designed MOF systems \cite{CCRodriguez2016, NATUREBaumann2019}, long-range interactions may break the site-permutation symmetries. {Indeed, for most of the lattices studied here, that is the case, with the exception of the kagome, the snub-hexagonal, and a variation of square-octagonal lattices. For these lattices, the strong constraint for the pNN coordination ($c_p$) is always fulfilled, leading to the form of $H_N$ being robust \cite{supmat}.} Therefore, the $\Gamma$-point degeneracies in the kagome lattice and the pseudospin-2 Dirac quasiparticle of the snub-hexagonal lattice are robust against long-range isotropic interactions. Particularly, for the snub-hexagonal, the space group predicts a twofold degeneracy, while the site-permutation symmetry shows that a fivefold symmetry is always preserved at $\Gamma$ \cite{PCCPCrasto2019}. {It is worth pointing out that even in the lattices fulfilling only $c_1$, the gap opened by the long-range coupling can be small, $\propto 2t_2$, where $t_2$ is the second-neighbor coupling. For instance, in graphene the ratio between second- and first-neighbor interaction is of only $\sim 4\%$ \cite{PRBReich2002}.}

\begin{figure}
\includegraphics[width=\columnwidth]{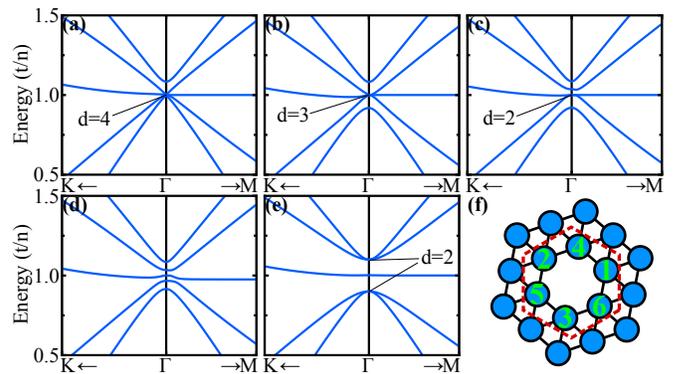}
\caption{\label{break-on-site} Band structure close to the $\Gamma$ point in the snub-hexagonal lattice with broken $P^{(1k)}$ site-permutation symmetries. (a) Lattice with breaking of $k=2$, (b)  $2 \le k \le 3$, (c) $2 \le k \le 4$, and (d) $2 \le k \le 5$ site-permutation symmetries. (e) Polar lattice with with broken $P^{(ij)}$ for $i=1,2$ and $3$ and $j=4,\,5$ and $6$. (f) Sublattice enumeration, with dashed line indicating the Wigner-Seitz cell.}
\end{figure}

\paragraph{Lattice perturbations.} Local perturbations in the lattice, e.g., site substitutions, or bond length distortions, will break the site-permutation symmetry. However, if not all site-permutation symmetries are broken, the Hamiltonian still commutes with a subset of these operations, yielding a smaller degenerate set of eigenvalues. 

To illustrate this behavior, we consider the snub-hexagonal lattice. Let us break specific site-permutation symmetries by changing the on-site energy of each site [Fig.\,\ref{break-on-site}(f)]. For instance, if all sites have identical energy, the $N=6$ Hamiltonian commutes with the five site-permutation symmetries $P^{(1k)}$, with $k=2,\,\dots,\,6$, leading to the fivefold degeneracy of Fig. \ref{tb-model}(e). By changing the on-site energy of site $k=2$, the Hamiltonian ceases to commute with $P^{(12)}$, but still commutes with the other four operations, leading to the fourfold degeneracy of Fig. \ref{break-on-site}(a). Similarly, by breaking two, three, and four of the site-permutation symmetries, a threefold, twofold, and nondegenerated points are seen in Figs. \ref{break-on-site}(b), \ref{break-on-site}(c), and \ref{break-on-site}(d), respectively. 

More generally, this shows that the site-permutation symmetries can generate HD points even in lattices with more than one type of site. The required criteria is that a subset of the site-permutation symmetries is still present. Considering that the sites split into two equivalent sets with $N_1$ and $N_2$ sites each, the Hamiltonian takes the form
\begin{equation}
	H_{N_1, N_2} = 
	\begin{pmatrix}
		H_{N_1} & v_0 V_{12} \\
		v_0 V_{12}^\dagger & H_{N_2}
	\end{pmatrix},
\end{equation}
where $H_{N_1}$ and $H_{N_2}$ are given by Eq. \eqref{hamilt} for the subset of site-permutation symmetries for each set of equivalent sites, $V_{12}$ is an all-ones rectangular $N_1 \times N_2$ matrix, and $v_0$ is the hopping intensity between the subsets. In this form $H_{N_1, N_2}$ will provide eigenvalues split into four sets: two sets of $N_1-1$ and $N_2-1$ degenerate states; and two other nondegenerate states.

For instance, let us consider splitting the snub-hexagonal lattice into a polar lattice with two nonequivalent sets of three sites, viz., sites $1$, $2$, and $3$ with on-site energy $\varepsilon_a$ and sites $4$, $5$, and $6$ with $\varepsilon_b \neq \varepsilon_a$. In this case we can still define two sets of permutation operations that commutes with the Hamiltonian, $P^{(1k)}$ with $k=2$, $3$ and $P^{(4l)}$ with $l=5$, $6$. These two sets allow for two double degenerated eigenvalues at the $\Gamma$ point, as shown in Fig. \ref{break-on-site}(e).

\paragraph{Conclusions.} In summary, {we have found weak and strong constraints in lattice design that yield HD eigenvalues in lattice Hamiltonians.} Lattices with $N$ equivalent sites in their unit cell, and azimuthal/spherical intersite interactions for 2D/3D systems, satisfying the coordination numbers criteria $c_1$ {(weak constraint)} or $c_p$ {(strong constraint)}, will present a Hamiltonian with a spectra consisting of two sets of eigenvalues: one nondegenerate state, and $N-1$ degenerate states. {The HD points in lattices fulfilling the strong constraint are robust against long-range interaction.
On the other hand, for the weak constraint a small gap is allowed and defined by the next NN hoppings. Nevertheless, the weak constraint allows for a richer number of lattices to be constructed. Interestingly, these emergent} HD points are not predicted by the lattice space groups, rather they are stabilized by a set of site-permutation symmetries. This feature is satisfied in the 2D and 3D lattices shown here. For the case of electronic systems, higher pseudospin Dirac fermions were shown to emerge in these HD points. More importantly, the generality of the discussion allows the interpretation of the results and lattice designs among different quasiparticle systems, e.g., electronic, phononic, photonic, magnonic, and cold-atom. {Our discussion has been focused on the $\Gamma$ point where the Bloch phases are zero, and the Hamiltonian could be interpreted as a graph theory adjacency matrix. A generalization for other $k$-points is a challenging task, since the site permutations do not correspond to geometrical transformations. Nonetheless, it is an interesting topic that still requires a breakthrough to expand our results for the whole Brillouin zone.}

\vspace{0.2cm}
\paragraph{Acknowledgments.}
The authors thank R. H. Miwa for valuable discussions. We acknowledge financial support from the Brazilian agencies CNPq, CAPES, and FAPEMIG, and the computational centers CENAPAD-SP and Laborat{\'o}rio Nacional de Computa{\c{c}}{\~a}o Cient{\'i}fica (LNCC-SCAFMat) for the computer time. 

\bibliography{bib}

\pagebreak
\widetext
\begin{center}
\textbf{\large Supplemental Materials: High-degeneracy points protected by site-permutation symmetries}

F. Crasto de Lima and G. J. Ferreira
\end{center}

\setcounter{equation}{0}
\setcounter{figure}{0}
\setcounter{table}{0}
\setcounter{page}{1}
\makeatletter
\renewcommand{\theequation}{S\arabic{equation}}
\renewcommand{\thefigure}{S\arabic{figure}}
\renewcommand{\bibnumfmt}[1]{[S#1]}
\renewcommand{ \citenumfont}[1]{S#1}

\section{HD Lattice construction}
\begin{figure}[h!]
\includegraphics[width=0.8\columnwidth]{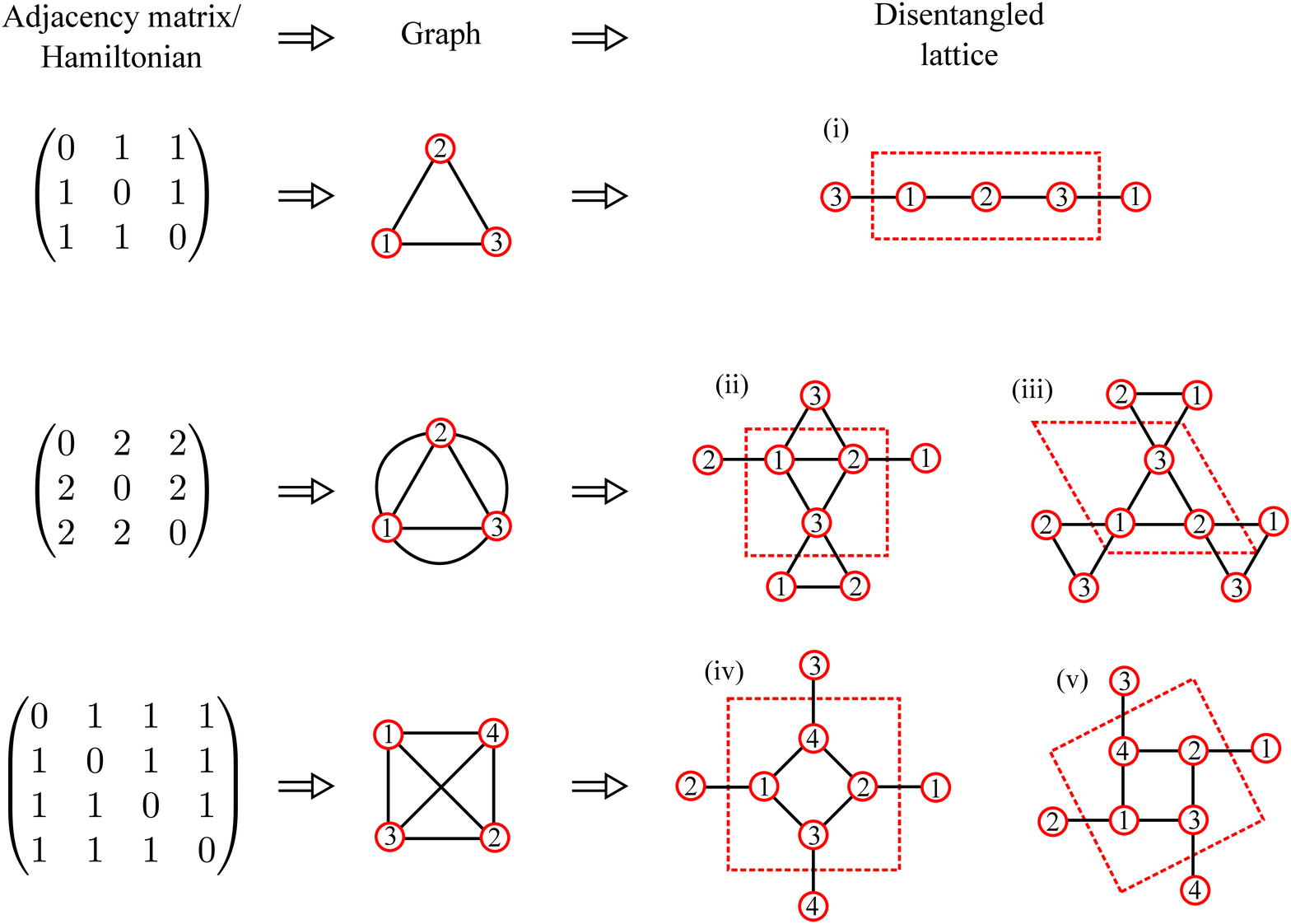}
\caption{\label{graph} Relation of the HD Hamiltonian, graph representation and disentangled graph in a periodic lattice.}
\end{figure}

The constructed, permutation symmetry commutating Hamiltonian has a structure of a graph adjacency matrix. The adjacency matrix is a square matrix, for which the non-diagonal terms indicate if two sites are adjacent, defining lines connecting such sites [main text Ref 38]. The diagonal terms indicate a site coupling with it self, which is represented by a loop. For instance, in Fig.\,\ref{graph}, we show the adjacency matrix and the related graph. This allows a direct relation of the HD Hamiltonian and the real space connection between the sites. With the graph constructed, it is possible to disentangle into periodic lattices. For instance, the all-one three site adjacency matrix, is represented by a triangular graph, which can be disentangled in a 1D lattice Fig.\,\ref{graph}(i). Taking the same three site system, but with a matrix of twos, the graph are triangular but each site double connected. In this case, the graph could be disentangled in two lattices, the triangular-rectangular and kagome, as indicated in Fig.\,\ref{graph}(ii) and (iii) respectively. Furthermore, for a four site system, with all one adjacency matrix/Hamiltonian, the graph can be represented by a center crossed square graph, which we could disentangle in two different square lattices, the square-octagonal and non-edge-to-edge square lattices, show in Fig.\,\ref{graph} panels (iv) and (v), respectively. To the best of the authors knowledge, there is current no systematical algorithms to find all possible disentangled lattices. Here we have follow a intuitive approach for this procedure.

\begin{figure}[h!]
\includegraphics[width=0.7\columnwidth]{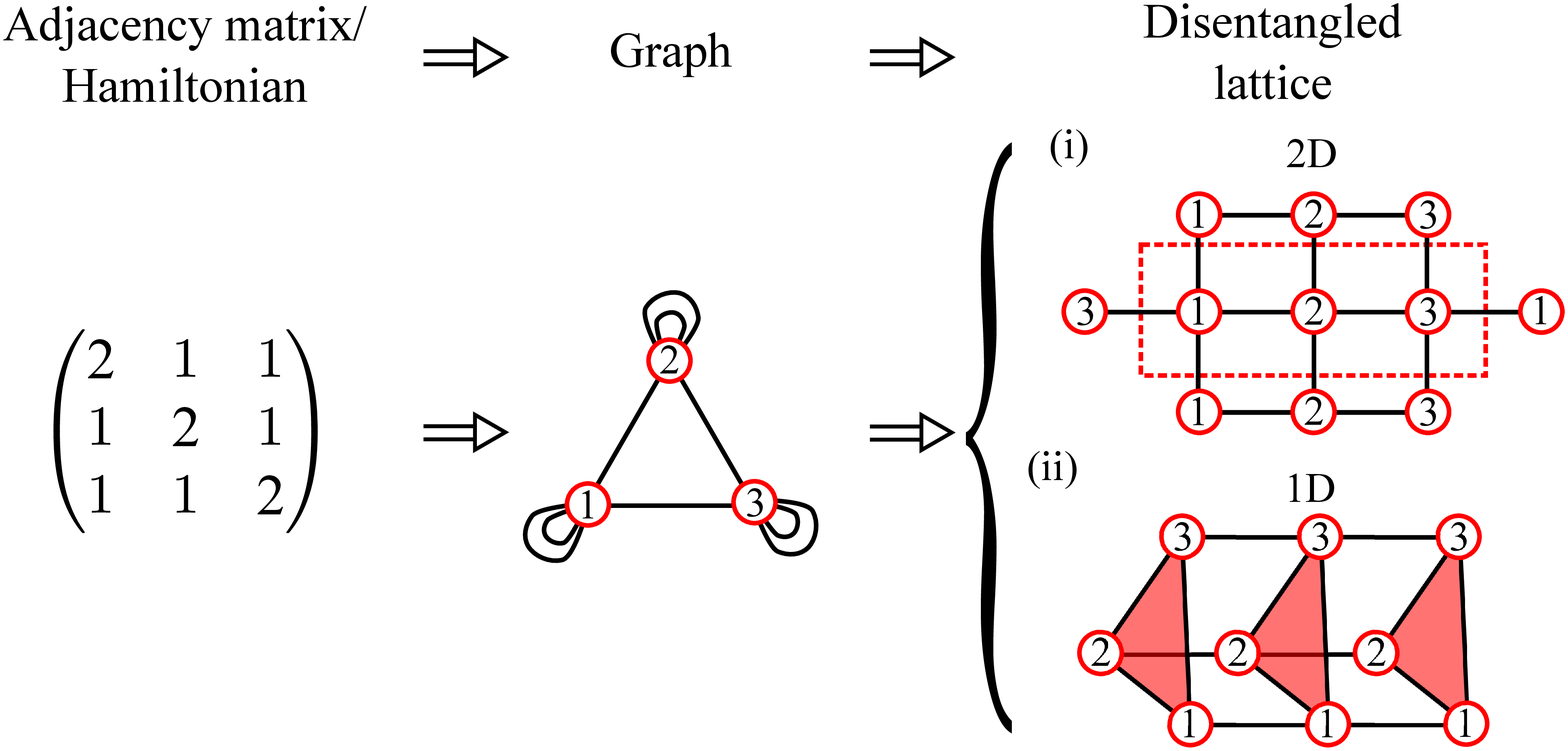}
\caption{\label{graph2} $3 \times 3$ adjacency matrix, graph representation and disentangled graph.}
\end{figure}

It is worth to mention that the same adjacency matrix can lead to lattices with different dimensionality. For instance, in Fig.\,\ref{graph2} we shown that the graph produced by a $3 \times 3$ adjacency matrix can be disentangled in a lattice periodic in 2D plane (i) and in a lattice periodic in a 1D line (ii).

\section{Long range coupling}
\begin{figure}[h!]
\includegraphics[width=0.9\columnwidth]{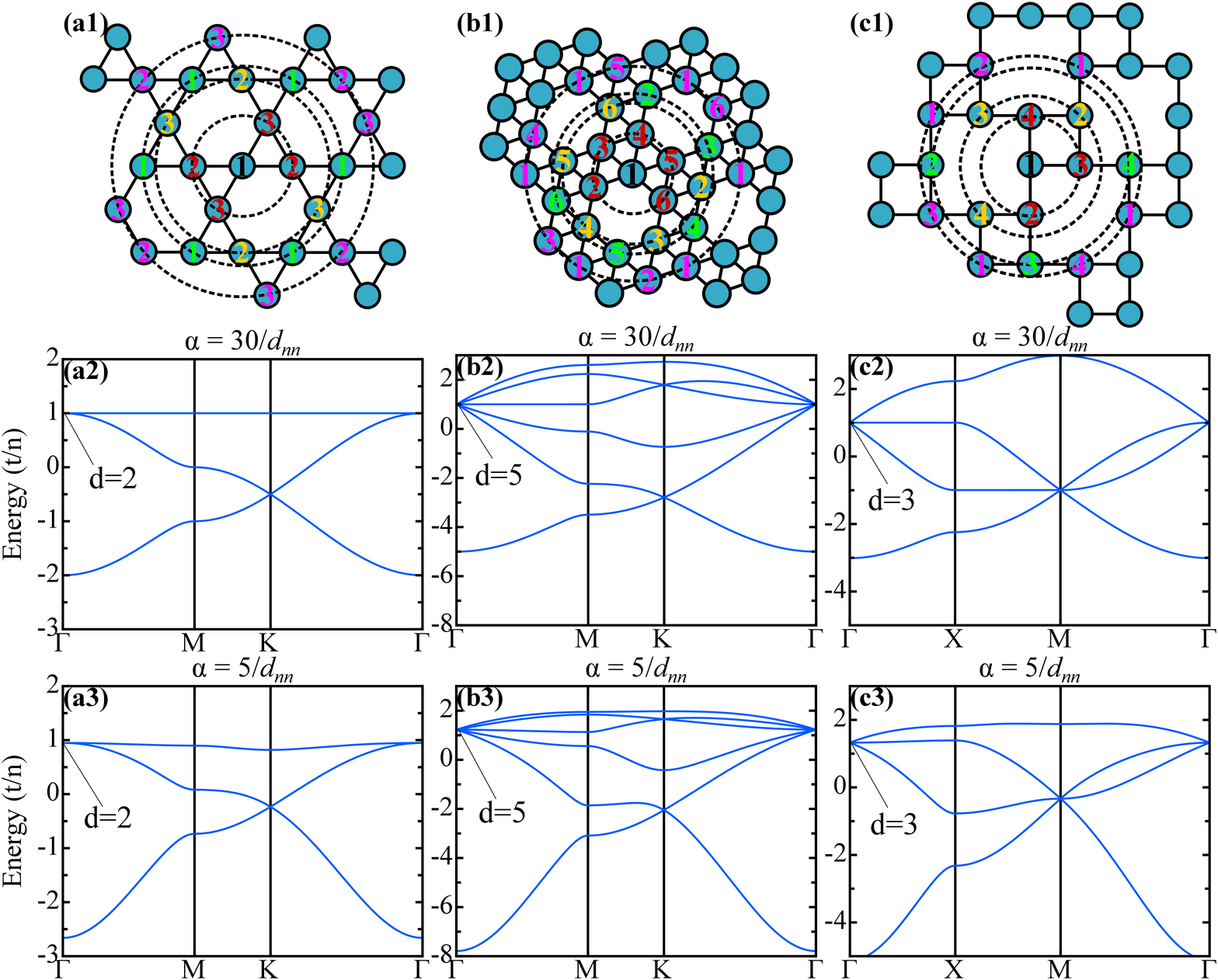}
\caption{\label{long-range} Atomic structure of kagome (a1), snub-hexagonal (b1) and non-edge-to-edge square (c1) lattices, where dashed circles indicate the different neighbor distance. TB model for nearest neighbor interaction $\alpha = 30/d_{\rm 1NN}$ (a2)-(c2), and long range interaction $\alpha = 5/d_{\rm 1NN}$ (a3)-(c3) for the kagome, snub-hexagonal and non-edge-to-edge square lattices, respectively.}
\end{figure}

Different from the discussed lattice within nearest neighbor interaction, long range interactions tends to couple further neighbor sites differently. Therefore, breaking the structure of the site-permutation symmetric Hamiltonian. However, within the discussed lattices, the kagome, snub-hexagonal and square non-edge-to-edge lattices are robust against long range interactions. Note that for these lattices further neighbors couplings, evidenced in Fig.\,\ref{long-range}(a1)-(c1), satisfy the $c_p$ criteria, with the Hamiltonian commuting with site-permutation symmetry. For instance, in the kagome lattice, within 1N coupling, the site $1$ connects two times with the site $2$ and $3$, marked in red, that accounts for a $c_1 = 4$, $n_1 = 2$ and $m_1 = 0$. In turn as the 1N Hamiltonian satisfy the $c_1$ criteria the system presents a twofold degeneracy, Fig.\,\ref{long-range}(a2). However, considering 2nd-nearest neighbor (2N) coupling, sites marked in yellow in Fig.\,\ref{long-range}(a1), the 2N criteria is satisfied $c_2 = 4$, $n_2 = 2$ and $m_2 = 4$. Furthermore for 3N and 4N couplings, sites marked in green and magenta, respectively, the criteria remains satisfied, for instance with $c_3 = 6$, $n_3 = 0$, and $m_3 = 6$, and $c_4 = 8$, $n_4 = 4$ and $m_4=0$. Therefore, the kagome lattice for long-range couplings keep the site-permutations symmetries in the Hamiltonian, consequently with its degeneracy preserved, as shown in Fig.\,\ref{long-range}(a3). A similar discussion can be made for the snub-hexagonal and square non-edge-to-edge lattices, where its 1N, 2N, 3N and 4N couplings, with sites marked in red, yellow, green and magenta, respectively, satisfying the $c_p$ criteria. That leads to the preserved degeneracy with long-range interaction, compare for instance Fig.\,\ref{long-range}(b2)-(c2) with (b3)-(c3).

\newpage

\section{Lattices geometries}
In this section we specify the lattice structure.

\subsection{kagome}
Hexagonal structure with P6/mmm space group, which assure at most twofold degeneracy at the $\Gamma$ point. The lattice can be described by the lattice vectors
\begin{eqnarray}
{\bm a}_1 &=& a \hat{\bm x}, \\
{\bm a}_2 &=& \frac{-a}{2} \hat{\bm x} + \frac{a \sqrt{3}}{2} \hat{\bm y}, \\
{\bm a}_3 &=& c \hat{\bm z},
\end{eqnarray}
and 3 sites given by the Wyckoff position g with coordinates
\begin{eqnarray}
{\bm r}_1 &=& \frac{1}{2} {\bm a}_1, \\
{\bm r}_2 &=& \frac{1}{2} {\bm a}_2, \\
{\bm r}_3 &=& \frac{1}{2} \left({\bm a}_1 + {\bm a}_2\right).
\end{eqnarray}

\subsection{snub-hexagonal}
Hexagonal structure with P6/m space group, which assure at most twofold degeneracy at the $\Gamma$ point. The lattice can be described by the lattice vectors
\begin{eqnarray}
{\bm a}_1 &=& a \hat{\bm x}, \\
{\bm a}_2 &=& \frac{-a}{2} \hat{\bm x} + \frac{a \sqrt{3}}{2} \hat{\bm y}, \\
{\bm a}_3 &=& c \hat{\bm z},
\end{eqnarray}
and 6 sites given by the Wyckoff position k with coordinates
\begin{eqnarray}
{\bm r}_1 &=& \frac{3}{7} {\bm a}_1 + \frac{1}{7} {\bm a}_2, \\
{\bm r}_2 &=& \frac{6}{7} {\bm a}_1 + \frac{2}{7} {\bm a}_2, \\
{\bm r}_3 &=& \frac{5}{7} {\bm a}_1 + \frac{4}{7} {\bm a}_2, \\
{\bm r}_4 &=& \frac{2}{7} {\bm a}_1 + \frac{3}{7} {\bm a}_2, \\
{\bm r}_5 &=& \frac{4}{7} {\bm a}_1 + \frac{6}{7} {\bm a}_2, \\
{\bm r}_6 &=& \frac{1}{7} {\bm a}_1 + \frac{5}{7} {\bm a}_2.
\end{eqnarray}

\subsection{triangular-rectangular}
Rectangular structure with Pmmm space group, which do not assure any degeneracy at the $\Gamma$ point. The lattice can be described by the lattice vectors
\begin{eqnarray}
{\bm a}_1 &=& a \hat{\bm x}, \\
{\bm a}_2 &=& \frac{a\sqrt{3}}{2} \hat{\bm y}, \\
{\bm a}_3 &=& c \hat{\bm z},
\end{eqnarray}
and 3 sites given by the Wyckoff positions n and h with coordinates
\begin{eqnarray}
{\bm r}_1 &=& 0, \\
{\bm r}_2 &=& \frac{1}{2} {\bm a}_1, \\
{\bm r}_3 &=& \frac{1}{4} {\bm a}_1 + \frac{1}{2}{\bm a}_2.
\end{eqnarray}

\subsection{square-octagonal}
Square structure with P4/mmm space group, which assure at most twofold degeneracy at the $\Gamma$ point. The lattice can be described by the lattice vectors
\begin{eqnarray}
{\bm a}_1 &=& a \hat{\bm x}, \\
{\bm a}_2 &=& a \hat{\bm y}, \\
{\bm a}_3 &=& c \hat{\bm z},
\end{eqnarray}
and 4 sites given by the Wyckoff position m with coordinates
\begin{eqnarray}
{\bm r}_1 &=& \frac{1}{2(\sqrt{2} +1)} {\bm a}_1 + \frac{1}{2}{\bm a}_2, \\
{\bm r}_2 &=& \frac{1}{2}{\bm a}_1 + \frac{1}{2(\sqrt{2} + 1)}{\bm a}_2, \\
{\bm r}_3 &=& \frac{1}{2}{\bm a}_1 + \frac{2\sqrt{2} + 1}{2(\sqrt{2} + 1)}{\bm a}_2, \\
{\bm r}_4 &=& \frac{2\sqrt{2} + 1}{2(\sqrt{2} + 1)}{\bm a}_1 + \frac{1}{2}{\bm a}_2.
\end{eqnarray}

\subsection{pyramidal}
Square structure with P4mm space group, which assure at most twofold degeneracy at the $\Gamma$ point. The lattice can be described by the lattice vectors
\begin{eqnarray}
{\bm a}_1 &=& a \hat{\bm x}, \\
{\bm a}_2 &=& a \hat{\bm y}, \\
{\bm a}_3 &=& c \hat{\bm z},
\end{eqnarray}
and 5 sites given by the Wyckoff position e and a with coordinates
\begin{eqnarray}
{\bm r}_1 &=& \frac{1}{2(\sqrt{2} +1)} {\bm a}_1 + \frac{1}{2}{\bm a}_2, \\
{\bm r}_2 &=& \frac{1}{2}{\bm a}_1 + \frac{1}{2(\sqrt{2} + 1)}{\bm a}_2, \\
{\bm r}_3 &=& \frac{1}{2}{\bm a}_1 + \frac{2\sqrt{2} + 1}{2(\sqrt{2} + 1)}{\bm a}_2, \\
{\bm r}_4 &=& \frac{2\sqrt{2} + 1}{2(\sqrt{2} + 1)}{\bm a}_1 + \frac{1}{2}{\bm a}_2, \\
{\bm r}_5 &=& \frac{1}{2} {\bm a}_1 + \frac{1}{2} {\bm a}_2 + \frac{2\sqrt{2} + 1}{2(\sqrt{2}+1)} \frac{a}{c}{\bm a}_3.
\end{eqnarray}

\subsection{cubic-octahedra}
Cubic structure with Pm$\bar{3}$m space group, which assure at most threefold degeneracy at the $\Gamma$ point. The lattice can be described by the lattice vectors
\begin{eqnarray}
{\bm a}_1 &=& a \hat{\bm x}, \\
{\bm a}_2 &=& a \hat{\bm y}, \\
{\bm a}_3 &=& a \hat{\bm z},
\end{eqnarray}
and 6 sites given by the Wyckoff position e with coordinates
\begin{eqnarray}
{\bm r}_1 &=& \frac{1}{2(\sqrt{2} +1)} {\bm a}_1 + \frac{1}{2}{\bm a}_2 + \frac{1}{2} {\bm a}_3, \\
{\bm r}_2 &=& \frac{1}{2}{\bm a}_1 + \frac{1}{2(\sqrt{2} + 1)}{\bm a}_2 + \frac{1}{2} {\bm a}_3, \\
{\bm r}_3 &=& \frac{1}{2}{\bm a}_1 + \frac{2\sqrt{2} + 1}{2(\sqrt{2} + 1)}{\bm a}_2 + \frac{1}{2} {\bm a}_3, \\
{\bm r}_4 &=& \frac{2\sqrt{2} + 1}{2(\sqrt{2} + 1)}{\bm a}_1 + \frac{1}{2}{\bm a}_2 + \frac{1}{2} {\bm a}_3, \\
{\bm r}_5 &=& \frac{1}{2} {\bm a}_1 + \frac{1}{2} {\bm a}_2 + \frac{2\sqrt{2} + 1}{2(\sqrt{2}+1)} {\bm a}_3, \\
{\bm r}_6 &=& \frac{1}{2} {\bm a}_1 + \frac{1}{2} {\bm a}_2 + \frac{1}{2(\sqrt{2}+1)} {\bm a}_3.
\end{eqnarray}

\subsection{crossed-tetragonal}
Tetragonal structure with P4$_2$/mmc space group, which assure at most twofold degeneracy at the $\Gamma$ point. The lattice can be described by the lattice vectors
\begin{eqnarray}
{\bm a}_1 &=& a \hat{\bm x}, \\
{\bm a}_2 &=& a \hat{\bm y}, \\
{\bm a}_3 &=& \frac{a}{\sqrt{2}} \hat{\bm z},
\end{eqnarray}
and 4 sites given by the Wyckoff position m with coordinates
\begin{eqnarray}
{\bm r}_1 &=& \frac{1}{4} {\bm a}_1 + \frac{1}{2} {\bm a}_2 + \frac{1}{4} {\bm a}_3, \\
{\bm r}_2 &=& \frac{3}{4} {\bm a}_1 + \frac{1}{2} {\bm a}_2 + \frac{1}{4} {\bm a}_3, \\
{\bm r}_3 &=& \frac{1}{2} {\bm a}_1 + \frac{1}{4} {\bm a}_2 + \frac{3}{4} {\bm a}_3, \\
{\bm r}_4 &=& \frac{1}{2} {\bm a}_1 + \frac{3}{4} {\bm a}_2 + \frac{3}{4} {\bm a}_3.
\end{eqnarray}

\subsection{pyroclore}
Hexagonal structure with P$\bar{6}$m2 space group, which assure at most two fold degeneracy at the $\Gamma$ point. The lattice can be describe by the lattice vectors
\begin{eqnarray}
{\bm a}_1 &=& a \hat{\bm x}, \\
{\bm a}_2 &=& \frac{-a}{2} \hat{\bm x} + \frac{a \sqrt{3}}{2} \hat{\bm y}, \\
{\bm a}_3 &=& a\sqrt{\frac{2}{3}} \hat{\bm z},
\end{eqnarray}
and 4 sites given by the Wyckoff positions j and d with coordinates
\begin{eqnarray}
{\bm r}_1 &=& \frac{1}{2} {\bm a}_1, \\
{\bm r}_2 &=& \frac{1}{2} {\bm a}_2, \\
{\bm r}_3 &=& \frac{1}{2} {\bm a}_1 + \frac{1}{2} {\bm a}_2, \\
{\bm r}_4 &=& \frac{1}{3} {\bm a}_1 + \frac{2}{3} {\bm a}_2 + \frac{1}{2} {\bm a}_3.
\end{eqnarray}

\end{document}